\documentclass[aps,twocolumn,prl,article]{revtex4}

\usepackage{graphicx}

\usepackage{dcolumn}
\usepackage{amsmath,amssymb,graphicx,bm,epsfig}
\usepackage{epstopdf}
\usepackage{color}
\usepackage{tabularx}

\renewcommand{\a}{\alpha}

\begin{document}
\def\EF{$E_\textrm{F}$}
\def\cblue{\color{blue}} 
\def\cred{\color{red}}  
\def\cgr{\color{green}}
\def\cmag{\color{magenta}}

\draft \preprint{J. Kim $et~al.$}

\title {Topological Phase Transition 
in an Archetypal $f$-electron Correlated System: Ce}

\author{Junwon Kim$^1$}
\email[Co-first authors: They contributed equally.]{}
\author{Dongchoon Ryu$^1$}
\email[Co-first authors: They contributed equally.]{}
\author{Chang-Jong Kang$^{1}$}
\email[Present address:
Department of Physics and Astronomy, Rutgers University,
Piscataway, New Jersey, 08854, USA]{}
\author{Kyoo Kim$^{1,2}$}
\author{Hongchul Choi$^1$}
\email[Present address:
       Scuola Internazionale Superiore di Studi Avanzati 
       Trieste 34136, Italy]{}
\author{T.-S. Nam$^1$}
\author{B. I. Min$^1$}
\email[e-mail: ]{bimin@postech.ac.kr}
\affiliation{
$^1$Department of Physics, Pohang University of Science and Technology, 
Pohang 37673, Korea\\
$^2$MPPHC\_CPM, Pohang University of Science and Technology, 
Pohang 37673, Korea \\
}
\date{\today}

\begin{abstract} 
A typical $f$-electron Kondo lattice system
Ce exhibits the well-known isostructural transition, 
the so-called ${\gamma}$-${\a}$ transition, 
accompanied by an enormous volume collapse.
Most interestingly, 
we have discovered that a topological-phase transition
also takes place in elemental Ce, 
concurrently with the ${\gamma}$-${\a}$ transition.  
Based on the dynamical mean-field theory approach 
combined with density functional theory,
we have unravelled that the non-trivial topology in ${\a}$-Ce 
is driven by the $f$-$d$ band inversion, 
which arises from 
the formation of coherent $4f$ band around the Fermi level.
We captured the formation of the $4f$ quasi-particle band
that is responsible for the Lifshitz transition and 
the non-trivial Z$_2$ topology establishment
across the phase boundary. 
This discovery provides a concept of ``topology switch"
for topological Kondo systems.
The ``on" and ``off" switching knob in Ce 
is versatile in a sense that it is controlled
by available pressure ($\sim 1$ GPa) at room temperature.
\end{abstract}

\maketitle

\begin{figure}[b]
\begin{center}
\includegraphics[width=0.5\textwidth]{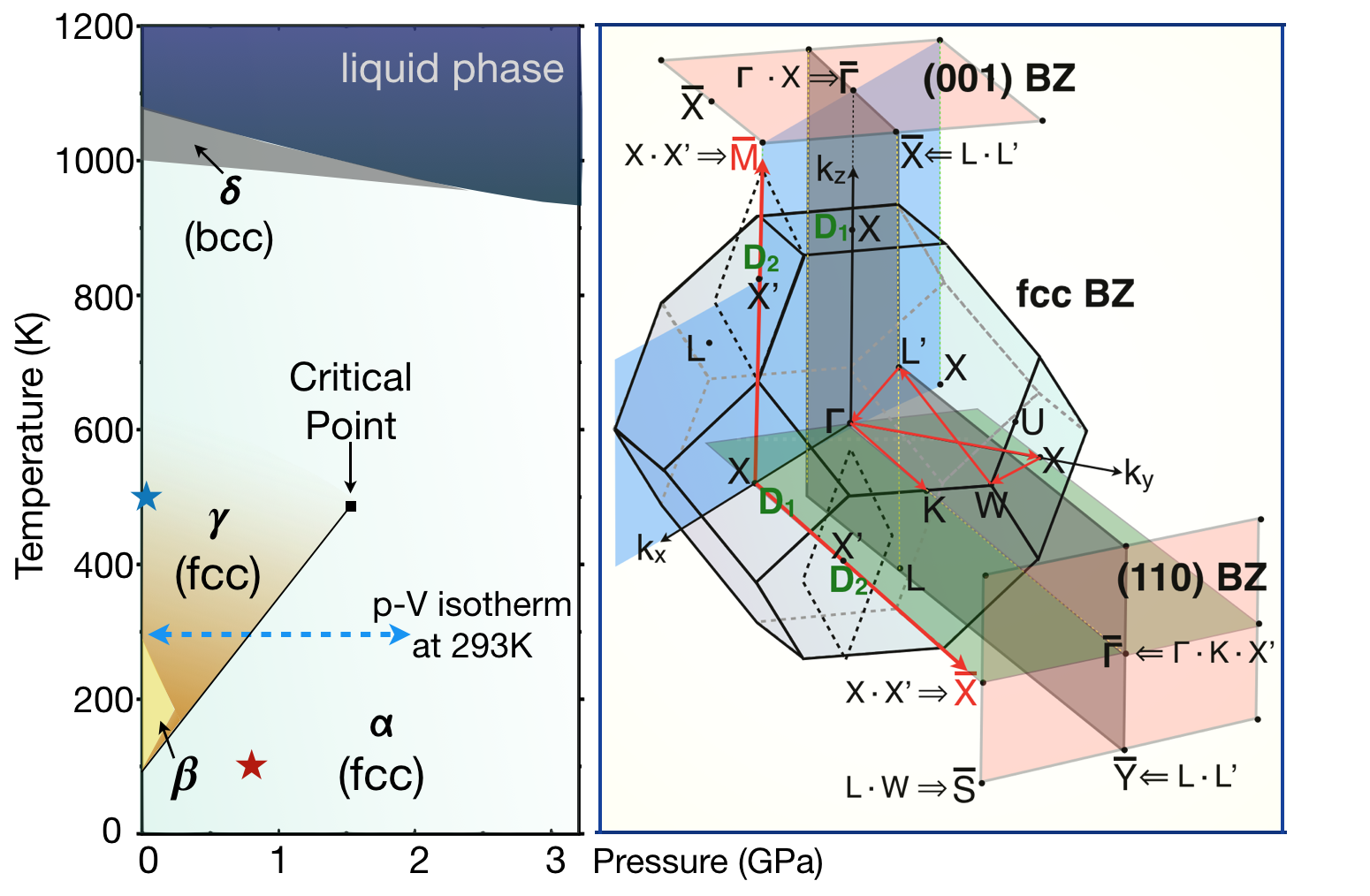}
\caption{
(Left) A phase diagram of Ce \cite{X-ray_diff}
(see also the supplement \cite{Supp}).
${\alpha}$-Ce at ``red star"
and  ${\gamma}$-Ce at ``blue star"
are selected for the comparison of electronic structures in
Fig.~\ref{basic_BAND}.
The blue-dotted line corresponds to the $P$-$V$ isotherm at 293 K. 
{\cred
}
(Right) The bulk BZ of fcc Ce and its (001) and (110) surface BZ.
There are two independent mirror planes of $k_{y}=0$ (in blue) 
and $k_{x}=k_{y}$ (in gray),
which, respectively, yield two mirror-symmetry lines along 
$\bar{M}$-$\bar{\Gamma}$-$\bar{M}$ and  
$\bar{X}$-$\bar{\Gamma}$-$\bar{X}$ in the (001) surface BZ.
Similarly, in the (110) surface BZ,
two mirror-symmetry lines are formed along
$\bar{Y}$-$\bar{\Gamma}$-$\bar{Y}$ and 
$\bar{X}$-$\bar{\Gamma}$-$\bar{X}$.
}
\label{phase_N_BZ}
\end{center}
\end{figure}

Physics of strongly-correlated $f$-electron 
materials has been a longstanding subject 
of special interest due to complex interplay
among the underlying interactions, such as strong Coulomb correlation,
spin-orbit (SO) coupling, and the hybridization of the 
localized $f$ and conduction electrons.  
More intriguing is that the interplay is very sensitive 
to small changes in the external parameters.
Elemental Ce, which has one occupied $f$ electron in its atomic phase,
is a prototypical $f$-electron Kondo lattice system 
exhibiting such sensitivity. 
Indeed Ce shows a rich phase diagram (see Fig. \ref{phase_N_BZ}) 
and many interesting physical properties
as a function of temperature ($T$) 
and pressure ($P$) \cite{mag_suscep,qc1,qc2,superconduct}.
The first-order isostructural volume-collapse transition 
from ${\gamma}$ to ${\alpha}$ phase of 
face-centered cubic (fcc) Ce
is the most representative phenomenon 
that experiences the sensitivity.
However, the driving mechanism of the ${\gamma}$-${\alpha}$ 
transition is still under debate,
between the two well-known models: Mott transition 
\cite{Mott_transition} vs. Kondo volume collapse 
\cite{Kondo_Volume_Collapse1}.
The current consensus is that 
there exists at least a significant change in the Kondo hybridization 
between the localized $4f$ electrons and conducting electrons 
across the transition 
\cite{Magnetic_form_factor,DMFT_f-local,dmft_opt_cond_Ce,Haule15}.
This peculiarity in Ce could facilitate the emergence of
nontrivial topology in the ground-state ${\alpha}$-phase of Ce.

\begin{figure*}[t]
\begin{center}
\includegraphics[width=0.89\textwidth]{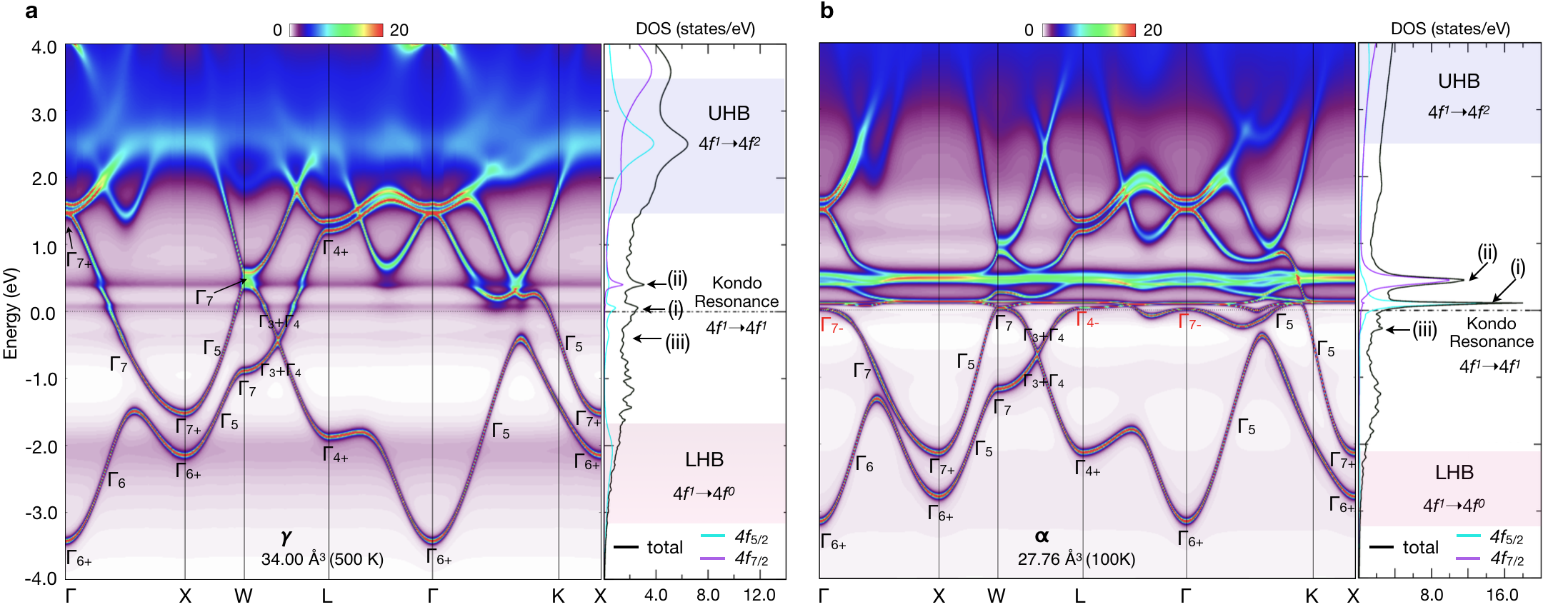}
\caption{
{\bf (a)} The DMFT electronic structure and DOS   
for ${\gamma}$-Ce calculated
at $V$ = 34 \AA$^3$ ($P$ = ambient pressure) and $T$ = 500 K,
and {\bf (b)} those for ${\alpha}$-Ce calculated
at $V$ = 27.76 \AA$^3$  ($P$ = 0.88GPa) and $T$ = 100 K.
The $4f$ spectral weights of both phases consist of mainly three parts:  
UHB at $2\sim 4$ eV,
LHB at $-2.0\sim -2.5$ eV, and the Kondo resonance near {\EF}.
In addition to the Kondo resonance near {\EF} (i),
the SO side peaks (ii), (iii) are seen at $\sim \pm 0.3$ eV.
Note that only ${\alpha}$-Ce shows the coherent quasi-particle
$4f$ band around {\EF},
which is shown more clearly in Fig. \ref{dmft_EF}.
}
\label{basic_BAND}
\end{center}
\end{figure*}
\begin{figure*}[t]
\begin{center}
\includegraphics[width=0.8\textwidth]{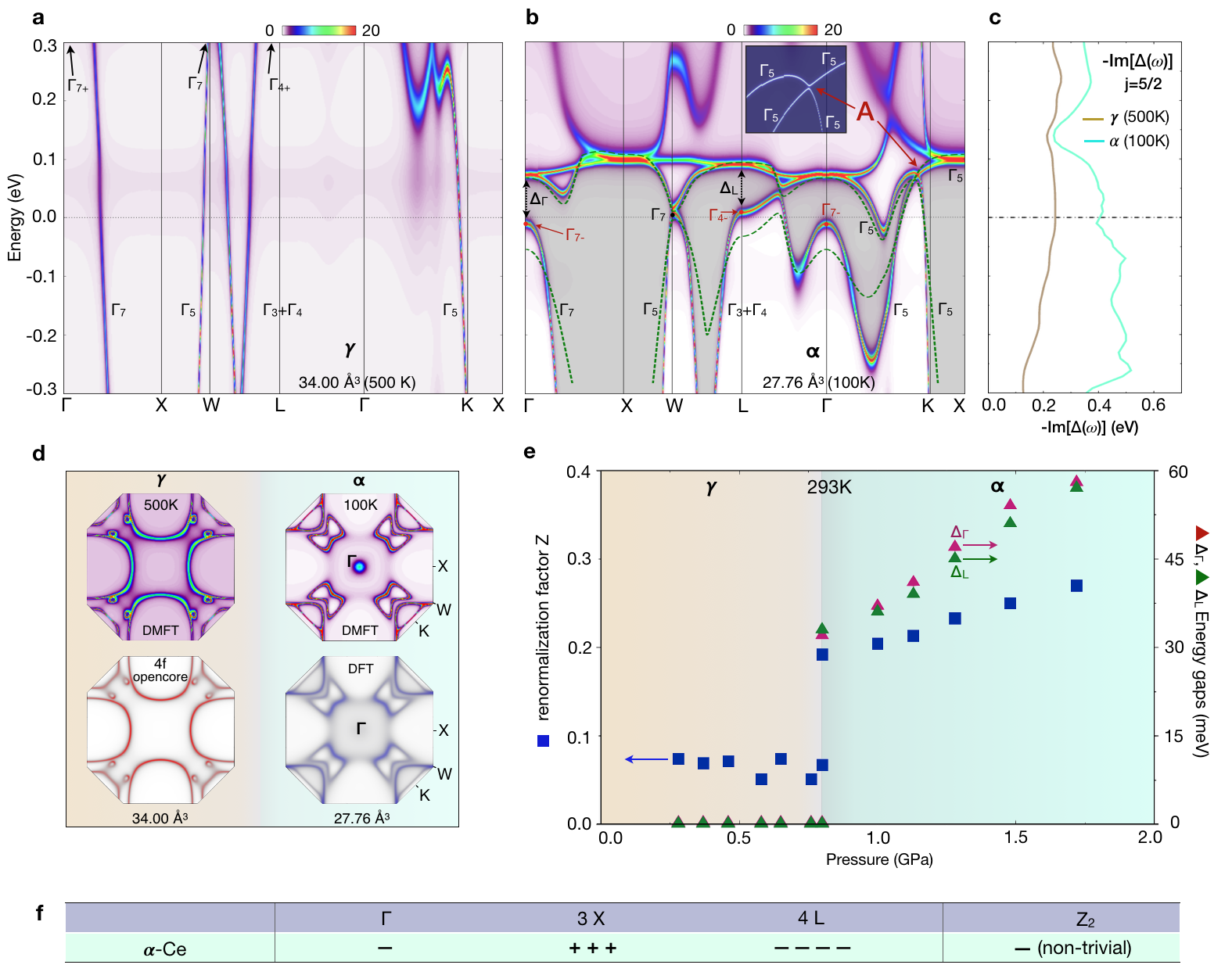} 
\caption{
The amplified DMFT electronic structures near {\EF}:
{\bf (a)} for ${\gamma}$-Ce and {\bf (b)} for  ${\alpha}$-Ce.
{\cred
}
For ${\gamma}$-Ce, $4f$ states are hardly seen, because they are incoherent.
For  ${\alpha}$-Ce, the coherent $4f$ bands formed around {\EF}
produce, via the hybridization with the conduction band,
the separated bands with the gap in-between (colored in gray).
There exist clear energy gaps at the TRIM points of ${\Gamma}$,  $X$ and $L$,
and also small energy gaps at $W$, in-between $L$-${\Gamma}$, 
and at ``A" along ${\Gamma}$-$K$.
The inset shows the gap formation at ``A",
arising from the same ${\Gamma}_5$ symmetry of 
the crossing bands \cite{Imsigma}.
The green-dotted lines overlaid with DMFT bands are the DFT bands 
rescaled by 1/2.
{\bf (c)} Imaginary part of the DMFT hybridization function $\Delta(\omega)$.
{\bf (d)} DMFT and DFT FSs for both phases (see also Fig. S1 \cite{Supp}).
{\bf (e)} The renormalization factor $Z$ 
and the energy gaps at ${\Gamma}$ and $L$ 
($\Delta_{\Gamma}$ and $\Delta_{L}$) 
are displayed as a function of pressure (see also Figs. S2-S4 \cite{Supp}).
The first-order-type phase transition is manifested
across the ${\gamma}$-${\a}$ transition.
{\bf (f)} The product of the parity eigenvalues 
of $\alpha$-Ce at 8 TRIM points in the fcc BZ.
{\cred
}
}
\label{dmft_EF}
\end{center}
\end{figure*}

In a recent theoretical work 
on topological Kondo insulator \cite{Coleman_KTI}, 
it is shown that
the Kondo hybridization in $f$-electron systems
can play an important role in 
the formation of non-trivial topology.
Since then, many subsequent studies have been reported 
to search for non-trivial topological materials, 
where the Kondo hybridization gap exists, {\it e.g.}, 
CeNiSn, CeRu$_4$Sn$_6$, Ce$_3$Bi$_4$Pt$_3$, SmB$_6$, SmS, 
and YbB$_{12}$
\cite{Coleman_KTI, CeRu4Sn6, Ce3Bi4Pt3, Fu13, Neupane13, SmS, YbB12}.
Despite extensive studies, however,
the topological nature of mother elements,
Ce, Sm, and Yb, supplying the correlated $f$-electrons 
to the above Kondo insulator compounds,
has not been explored yet.  Here we report,
based on the dynamical mean-field theory (DMFT) approach 
combined with density functional theory (DFT)
that has been successful to describe
electronic structures of Ce and Ce compounds
\cite{dmft_opt_cond_Ce,DMFT_f-local,CeIrIn5,CePd3},
that a narrow $f$-band metal ${\alpha}$-Ce 
has the non-trivial topology of topological-insulator (TI)-type and 
topological-crystalline-insulator (TCI)-type nature, and 
the topological phase transition and the Lifshitz electronic 
transition occur concomitantly with the ${\gamma}$-${\a}$
volume collapse transition in Ce.

Figure  \ref{basic_BAND} shows the DMFT band structures 
and the densities of states (DOSs) of 
${\gamma}$- and ${\alpha}$-Ce.
In the DMFT calculations, we have used the Coulomb correlation ($U$) 
and the exchange ($J$) interaction parameters 
of $U=5.5$ eV and $J=0.68$ eV for the Ce $f$-electrons
(refer to the supplement for the computational details) \cite{Supp}.
The $4f$ spectral weights of both phases 
have three main parts in common:
the lower Hubbard band (LHB) at $-2.0 \sim -2.5$ eV
corresponding to the $4f^0$ final state,
the upper Hubbard band (UHB) at $2 \sim 4$ eV
corresponding to the $4f^2$ final state,
and the Kondo resonances near the Fermi level ({\EF})
corresponding to the $4f^1$ final states.
{\cred
}
The energy positions of LHB and UHB are in good agreement 
with photoemission spectroscopy (PES) \cite{PES0,PES1,resonant_PES} 
and inverse PES experiments \cite{inverse_PES}.
{\cred
}
One of the most notable features in Fig. \ref{basic_BAND} is that
the spectral weight of the Kondo resonance
around {\EF} is much stronger in ${\alpha}$-Ce than in ${\gamma}$-Ce,
and exhibits the coherent quasi-particle band feature in ${\alpha}$-Ce,
as is consistent with previous PES 
\cite{PES0,PES1,resonant_PES,Chen18,inverse_PES} 
and theoretical reports 
\cite{Ce-dmft1,Ce-dmft2,Amadon15,Ce-dmft3}.
As will be discussed below, 
these contrasting Kondo-resonance features between 
the two phases lead to the quite different topological classes:
trivial and non-trivial Z$_2$ topologies for ${\gamma}$-Ce and
${\alpha}$-Ce, respectively. 
 

\begin{figure*}[t]
\begin{center}
\includegraphics[width=0.80\textwidth] {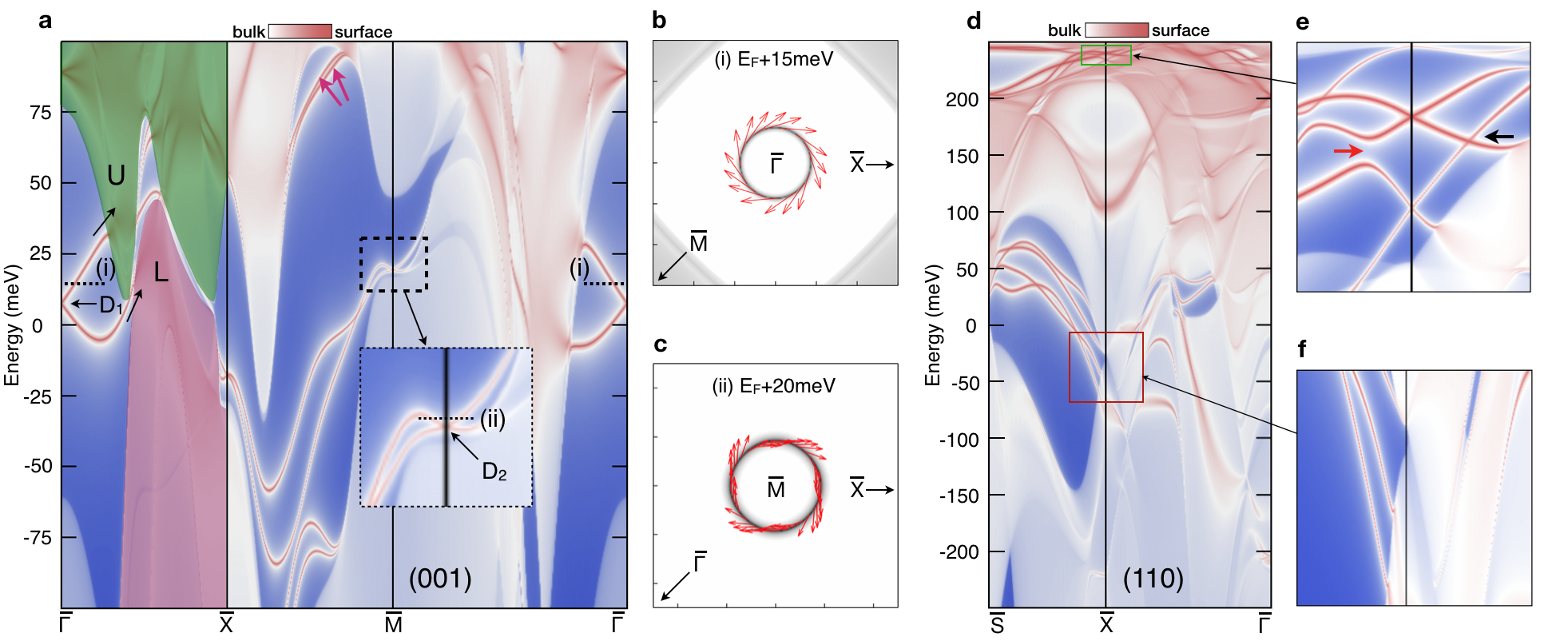} 
\caption{
(Color online) 
{\bf (a)} The (001) surface electronic structure of ${\alpha}$-Ce, 
calculated by the tight-binding (TB) model with semi-infinite slabs.
The TB Hamiltonian is constructed from the DFT band result 
(rescaled by 1/2 near {\EF}).
{\cred
}
{\bf (b),(c)} The helical spin structures of the 
``D$_1$" and ``D$_2$" Dirac-cone energy surfaces,
as indicated by (i) and (ii), respectively.
{\bf (d)} The (110) surface electronic structure of ${\alpha}$-Ce.
{\bf (e),(f)} Amplified band structures inside the green-square
and the red-square, respectively, in (D). 
In (E), TSSs of a typical TCI-type nature are revealed 
with the gapped (red arrow) and protected (black arrow) Dirac points,
while, in (F), TSSs are mostly buried
under the bulk-projected bands.
}
\label{TSS_band}
\end{center}
\end{figure*}


The incoherent and coherent $4f$ spectral weights 
for ${\gamma}$-Ce and ${\alpha}$-Ce, respectively,
are more clearly shown in the amplified DMFT 
band structures in Fig. \ref{dmft_EF}.
It is seen in Fig. \ref{dmft_EF}a that, 
for ${\gamma}$-Ce, $4f$ electrons are incoherent, and so 
mainly the $5d$ band crosses {\EF},
which agrees well with the optical spectroscopy result \cite{opt_spec_Ce}.
In contrast, for ${\a}$-Ce, 
the coherent 4$f$ quasi-particle band feature
is evident near {\EF} in Fig. \ref{dmft_EF}b,
which is the origin of the effective-mass enhancement 
of charge carriers and
the change of the charge carrier character from $5d$ to $4f$.
The coherent band feature for ${\a}$-Ce is corroborated by the fact that 
the DMFT bands have almost the same dispersion 
as the renormalized DFT bands rescaled approximately by 1/2 
(dotted green-line in Fig. \ref{dmft_EF}b).

The different electronic structures between the two phases are 
also reflected in the Fermi surfaces (FSs).
The shapes of FSs in Fig. \ref{dmft_EF}d 
are topologically different, suggesting that 
${\gamma}$-${\alpha}$ transition 
{\cred
}
corresponds to the Lifshitz transition (see the supplement) \cite{Supp}.
It is noteworthy in Fig. \ref{dmft_EF}d
that, while the DMFT FS of ${\gamma}$-Ce is very close to
that obtained from the DFT-opencore (``4f-opencore") calculation
considering the $4f$ electrons as core electrons,
the DMFT FS of ${\a}$-Ce is quite similar to the DFT FS.
These results indicate that, for ${\gamma}$-Ce, 
the contribution of 4$f$ electrons to the FS is negligible,
and, for ${\a}$-Ce, the 4$f$ quasi-particle band at {\EF}
can be described properly by the DFT band
(see Fig. S1 of the supplement) \cite{Supp}.

The key ingredient that makes the difference has something to do with 
the degree of the renormalization factor (quasi-particle weight) $Z$, 
arising from the Coulomb correlation interaction of 4$f$ electrons.
The renormalization factor $Z$ is obtained from the slope of the
real-part of the self-energy $\Sigma(\omega)$ at {\EF}.
As shown Fig. S4 of the supplement \cite{Supp}, 
we have obtained qualitatively different behaviors 
of $\Sigma(\omega)$'s between the ${\a}$ and ${\gamma}$ 
phases, which produce quite distinct electronic structures
and resulting physical parameters.
Indeed, Fig. \ref{dmft_EF}e shows that 
{\cred
}
$Z$ increases discontinuously
across the ${\gamma}$-${\a}$ transition.
As a result,
both the hybridization strength ${\Delta}({\omega})$
(Fig. \ref{dmft_EF}c) 
and the $f$-$f$ hopping strength,
which are to be effectively proportional to $Z$,
are enhanced for ${\a}$-Ce,
which give rise to the enhanced 4$f$ 
spectral weight and help to form the coherent 4$f$ band 
around {\EF} \cite{Amadon15}.

The evolution of the electronic structure 
across the ${\gamma}$-${\alpha}$ transition 
makes the elemental Ce more interesting in a topological sense.
{\cred
}
The coherent quasi-particle band in ${\a}$-Ce,
which, via the hybridization with the conduction band, 
brings about the hybridization gap in the ${\a}$-Ce phase, 
as indicated by gray-shaded area in Fig. \ref{dmft_EF}b. 
The energy gaps are clearly seen
at every time-reversal invariant momentum (TRIM) points of
${\Gamma}$, $X$, and $L$,
while those at $W$ and in-between $L$-$\Gamma$
are barely gapped. 
{\cred
}
Then, with respect to the hybridization gap,
the 5$d$ band of even parity and the 4$f$ band of odd parity
are inverted at the TRIM point X. 
Since the crystal structure is symmetric 
under the inversion operation,
the additional odd parity to the TRIM points yields
the non-trivial Z$_2$ topology of ${\a}$-Ce, 
as shown in Fig. \ref{dmft_EF}f.

Figure \ref{dmft_EF}e shows that the non-trivial Z$_2$ topology 
in ${\a}$-Ce is established at the very starting edge 
(volume 28 \AA$^3$ at 293 K) of $\alpha$-phase
in the $\gamma$-${\a}$ transition.
Note that no gaps are present in the $\gamma$ phase,
but the gaps at the TRIM points, $\Delta_{\Gamma}$ and $\Delta_{L}$ 
of about 30 meV ($\Delta_{X} > 2$ eV), are suddenly developed 
in the $\alpha$ phase.
{\cred
}
This implies that
the topological phase transition
would occur concomitantly with 
the $\gamma$ to $\alpha$ volume collapse transition.
The more detailed evolution of the band structures 
across the $\gamma$-$\a$ transition is given 
along a $P$-$V$ isotherm at 293 K in Fig. S2 of the supplement \cite{Supp}.
{\cred
}

In order to confirm
the non-trivial Z$_2$ topological invariance of ${\a}$-Ce, 
we have performed the surface electronic structure calculations
for the slab geometry of ${\a}$-Ce with (001) surface
and explored the existence of topological surface states (TSSs)
in  Fig. \ref{TSS_band}a. 
Note that, as shown in Fig. \ref{phase_N_BZ}, 
one $X$ point is projected onto $\bar{\Gamma}$,
while two non-equivalent $X$ and $X'$ 
are projected onto $\bar{M}$ of the (001) surface BZ.
Indeed, as shown in Fig. \ref{TSS_band}a,
the TSSs and corresponding Dirac points emerge 
in the indirect gap region at $\bar {\Gamma}$ (``D$_1$")
and $\bar {M}$ (``D$_2$").
Due to the bulk metallic nature of ${\a}$-Ce, 
most part of the Dirac bands at $\bar {M}$ is buried under the 
bulk-projected bands and so the band connectivity is not clear.
Nevertheless, it is evident in Fig. \ref{TSS_band}a that 
the surface states along $\bar {\Gamma}$-$\bar{X}$ are 
the Dirac-cone states,
because the lower surface band reaches the violet-colored band  (``L")
below the indirect gap,
while the upper one reaches the green-colored band (``U")
above the indirect gap.
The helical spin texture of the corresponding Dirac cone FS 
around  $\bar {\Gamma}$ and $\bar {M}$
in Fig. \ref{TSS_band}b and \ref{TSS_band}c
also manifests the spin-momentum locking behavior,
reflecting its topological nature.

The double Dirac points, which are supposed to be at $\bar{M}$ 
due to the projection of two non-equivalent X TRIM points
(Fig. \ref{phase_N_BZ}),
are to be separated due to the hybridization between the bands
of the double Dirac cones.
{\cred
}
On the (001) surface Brillouin-Zone (BZ) of ${\a}$-Ce,
there are two mirror symmetric lines, $\bar{\Gamma}$-$\bar{X}$ 
and $\bar{\Gamma}$-$\bar{M}$, as shown in Fig. \ref{phase_N_BZ},
which could play a key role in realizing 
the TCI-type nature.
It is thus obvious that the band crossing along $\bar{X}$-$\bar{M}$
that is not a mirror symmetric line would be gapped, 
but that along $\bar{M}$-$\bar{\Gamma}$
needs further consideration.
However, the surface states along $\bar{M}$-$\bar{\Gamma}$ are
completely buried under the bulk projected bands,
and so it is not easy to identify the specific TCI-type band feature 
in Fig. \ref{TSS_band}a. 
In view of the surface states inside the dotted-black square
and those along $\bar{X}$-$\bar{M}$ designated by red arrows 
in Fig. \ref{TSS_band}a,
we just conjecture that the band crossing 
along $\bar{M}$-$\bar{\Gamma}$
would be gapped to have Rashba-type surface states, 
as reported for the golden phase of SmS ($g$-SmS) 
that is expected to have
the same topological symmetry as ${\a}$-Ce \cite{CJKang15}.
In fact, ${\a}$-Ce is found to have 
the same mirror Chern numbers as $g$-SmS \cite{CJKang2}, 
as shown in Fig. S5 of the supplement \cite{Supp}.

We have also examined 
the TSSs for the (110) and (111) surfaces of ${\a}$-Ce.  
For the (110) surface, single and double Dirac points are expected
to be located at $\bar{\Gamma}$ and $\bar{X}$, respectively, 
as shown in Fig. \ref{phase_N_BZ}. 
For the (111) surface, only the single Dirac point is expected
at $\bar{M}$, as shown in Fig. S6 of the supplement \cite{Supp}.
As shown in Fig. \ref{TSS_band}d and Fig. S6, however,
neither (110) nor (111) surface states show
a clear TI-type or TCI-type signature in the hybridization gap region,
because, here too, most of the surface states near {\EF} are buried 
under the bulk-projected bands.
In this circumstance, for the (110) surface,
one apparent TCI signature is seen at $\bar{X}$ near 240 meV 
in Fig. \ref{TSS_band}e,
which demonstrated the gapped and protected Dirac points
along  $\bar{X}$-$\bar{S}$ (red arrow) 
and $\bar{X}$-$\bar{\Gamma}$ (black arrow), respectively.
This suggests that 
the near-{\EF} TSSs buried under the bulk-projected bands
in  Fig. \ref{TSS_band}f would also have the TCI-type band nature.

Our finding highlights that 
a typical narrow $f$-band metal
${\alpha}$-Ce is a topological Kondo system of TI- 
and TCI-type nature, and the ``on" and ``off"
topology switch can be operative
by using a $P$-tuning or $T$-tuning knob,
accompanied by the first-order 
volume-collapse and Lifshitz transitions. 
So Ce would be an excellent test-bed for investigating 
the topological phase transition in 
$f$-electron Kondo lattice systems.
It is thus highly desirable to explore the topological 
surface states in ${\alpha}$-Ce, preferentially for its (001) surface,
by using high resolution angle-resolved PES measurement.

{\bf Acknowledgments}$-$
We would like to thank J.D. Denlinger, J.-S. Kang, and J.H. Shim 
for helpful discussions.
This work was supported by the NRF (Grant No. 2017R1A2B4005175,
Grant No. 2018R1A6A3A01013431, Grant No. 2016R1D1A1B02008461),
Max-Plank POSTECH/KOREA Research Initiative (No. 2016K1A4A4A01922028), 
the POSTECH BSRI Grant, 
and the KISTI supercomputing center (Grant No. KSC-2017-C3-0057).



\begin{thebibliography}{99}
\bibitem{Imsigma}
Here, to capture the quasi-particle feature,
the imaginary part of self-energy 
(Im${\Sigma}$(${\omega}$)) is set to be zero,
namely, the Hamiltonian is assumed to be Hermitian.
\end{thebibliography}
\end{document}


\def\EF{$E_\textrm{F}$}
\def\cblue{\color{blue}} 
\def\cred{\color{red}}  
\def\cgr{\color{green}}
\def\cmag{\color{magenta}}

\draft \preprint{J. Kim $et~al.$}

\title {Supplement of 
``Topological Phase Transition 
in an Archetypal $f$-electron Correlated System: Ce"
}

\author{Junwon Kim$^1$}
\email[Co-first authors: They contributed equally.]{}
\author{Dongchun Ryu$^1$}
\email[Co-first authors: They contributed equally.]{}
\author{Chang-Jong Kang$^{1}$}
\email[Present address:
Department of Physics and Astronomy, Rutgers University,
Piscataway, New Jersey, 08854, USA]{}
\author{Kyoo Kim$^{1,2}$}
\author{Hongchul Choi$^1$}
\email[Present address:
       Scuola Internazionale Superiore di Studi Avanzati 
       Trieste 34136, Italy]{}
\author{T.-S. Nam$^1$}
\author{B. I. Min$^1$}
\email[e-mail: ]{bimin@postech.ac.kr}
\affiliation{
$^1$Department of Physics, Pohang University of Science and Technology,
        Pohang 37673, Korea\\
$^2$MPPHC\_CPM, Pohang University of Science and Technology,
        Pohang 37673, Korea \\
}

\maketitle

\section{Computational details} 

{\bf DFT+DMFT:}
To describe the topological phase transition 
across the ${\gamma}$-${\alpha}$ transition,
it is important to have accurate band structures 
of both ${\alpha}$ and ${\gamma}$ phases 
since the band structures give the information on 
which phase might have the non-trivial Z${_2}$ topology.
Most of all, the correlation effect should carefully be considered 
since Ce $4f$ electrons are spatially much localized.
To consider the correlation effect properly,
we have used the charge self-consistent DFT+DMFT scheme \cite{DMFT}
implemented in the full-potential code of WIEN2k \cite{WIEN2k}.

In the actual DMFT problem, 
the solution of the auxiliary quantum impurity problem
is achieved by the continuous time quantum Monte Carlo (CTQMC) solver
\cite{CTQMC1,CTQMC2}.
For the double-counting functional,
we used the nominal double-counting scheme  
introduced in Ref. \cite{nominal_dc1,nominal_dc2}. 
The on-site Coulomb correlation ($U$) and the
exchange ($J$) interaction parameters 
were set by 5.5 eV and 0.68 eV, respectively,  
and the spin-orbit coupling is included.

{\bf Tight-binding model for surface states:}
From the comparison of DMFT and DFT band structures of ${\alpha}$-Ce,
we have found that the renormalized DFT band is useful 
to investigate the low-energy band structure,
and so investigated the surface electronic structures  
of ${\alpha}$-Ce,
using the tight-binding (TB) model constructed from
the renormalized DFT band result (rescaled by 1/2 at around EF).
For the extraction of TB model parameters,
the maximally localized Wannier functions were used 
\cite{Wannier1,Wannier2,Wannier3}.
Then we constructed the TB Hamiltonian employing the Wannier90 code 
\cite{Wannier_code}
and performed semi-infinite TB slab calculations 
to obtain the surface states of $\alpha$-Ce,
using the Green function scheme\cite{Green_function_scheme} 
implemented in Wannier tools \cite{Wannier_tools}.

\begin{figure*}[t]
\begin{center}
\includegraphics[width=0.95\textwidth]{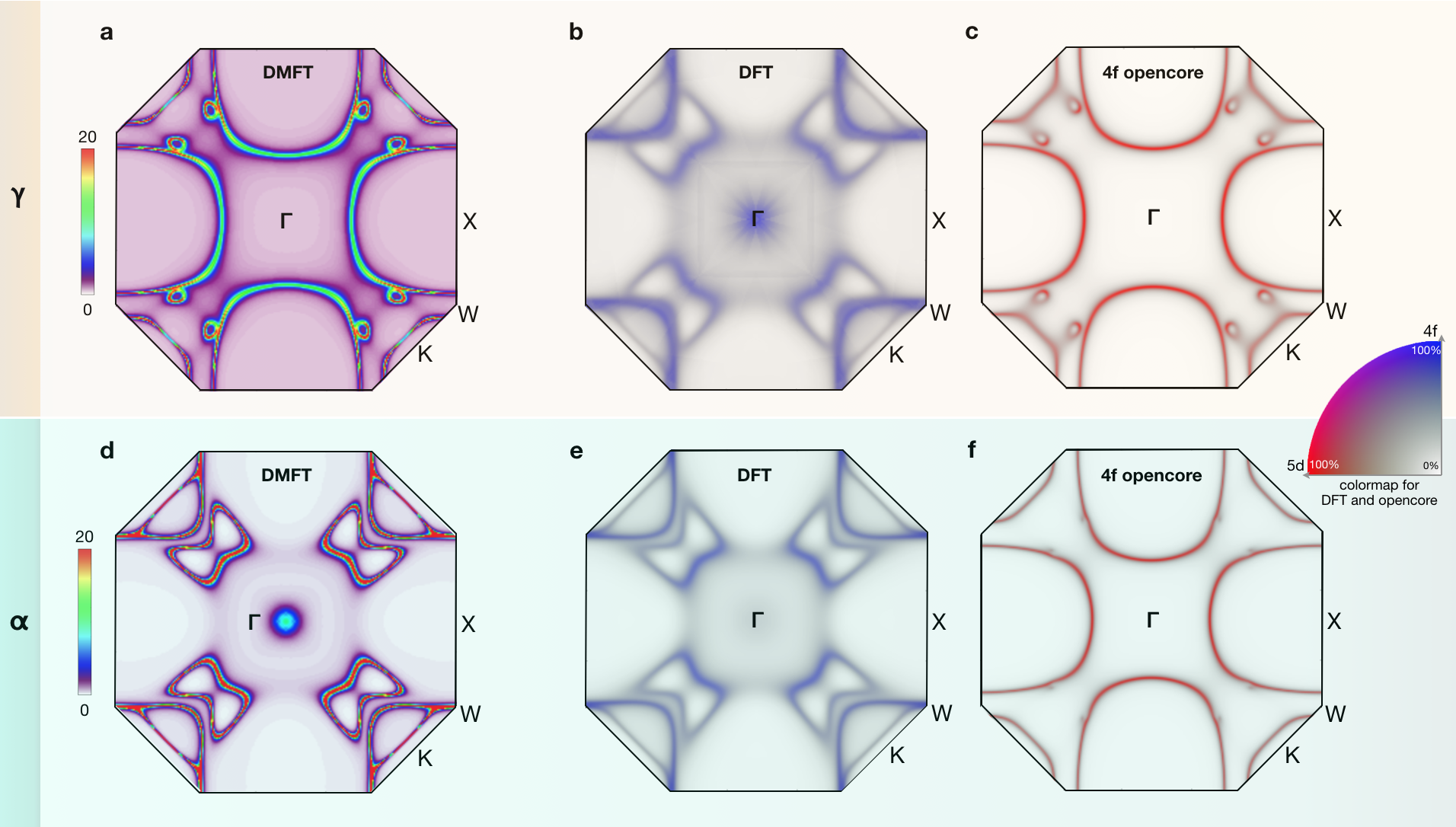} 
\caption{
{\bf (a)-(c)} 
Fermi surfaces (FSs) of ${\gamma}$-Ce calculated 
at $T$ = 500 K and $V$ = 34 \AA$^3$
by DMFT, DFT, and DFT-opencore (``4f-opencore") methods,
respectively.  
Note that the DMFT FS is very similar to the ``4f-opencore" FS,
which is composed of mostly $5d$ electrons.
{\bf (d)-(f)} FSs of ${\alpha}$-Ce calculated
at $T$ = 100 K and $V$ = 27.76 \AA$^3$
by DMFT, DFT, and DFT-opencore (``4f-opencore") 
methods, respectively.
The DMFT FS shows a very good agreement with the DFT FS,
which is composed of mostly $4f$ electrons.
The seemingly blurred FSs from the DFT and DFT-opencore calculations
originate just from the effect of the Gaussian broadening of each band.
}
\label{FS_compare}
\end{center}
\end{figure*}

\begin{figure}[t]
\begin{center}
\includegraphics[width=0.9\textwidth] {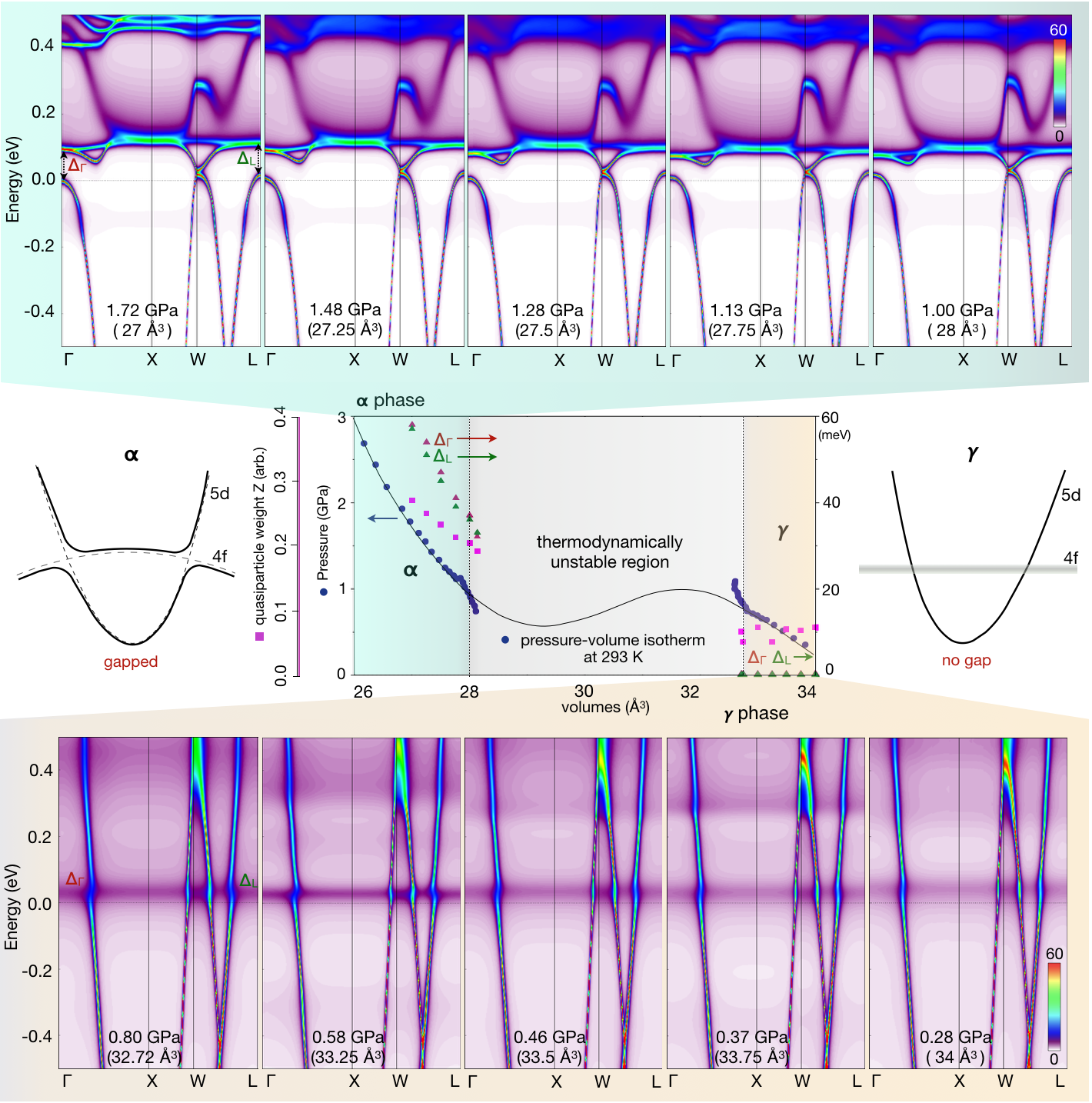} 
\caption{
(Color online) 
The pressure-volume isotherm at $T$ = 293 K \cite{X-ray_diffraction}
and the evolution of gaps $\Delta_{\Gamma}$ and $\Delta_{L}$ 
at ${\Gamma}$ and $L$ as a function of volume.
Maxwell construction is used to capture
the end volume of the ${\gamma}$ phase 
and the starting volume of the ${\a}$ phase,
which correspond to about $V$ = 32.75  \AA$^3$ and $V$ = 28 \AA$^3$, 
respectively.
At the very starting edge of the ${\alpha}$ phase, 
the gaps $\Delta_{\Gamma}$ and $\Delta_{L}$ become finite,
in contrast to almost zero gap feature in the ${\gamma}$ phase,
The DMFT band structures as a function of volume 
are also displayed, which clearly shows the coherent $4f$ 
band formation for ${\a}$-Ce with volume less than 28 \AA$^3$.
}
\label{gap_feature}
\end{center}
\end{figure}

\begin{figure}[t]
\begin{center}
\includegraphics[width=0.99\textwidth] {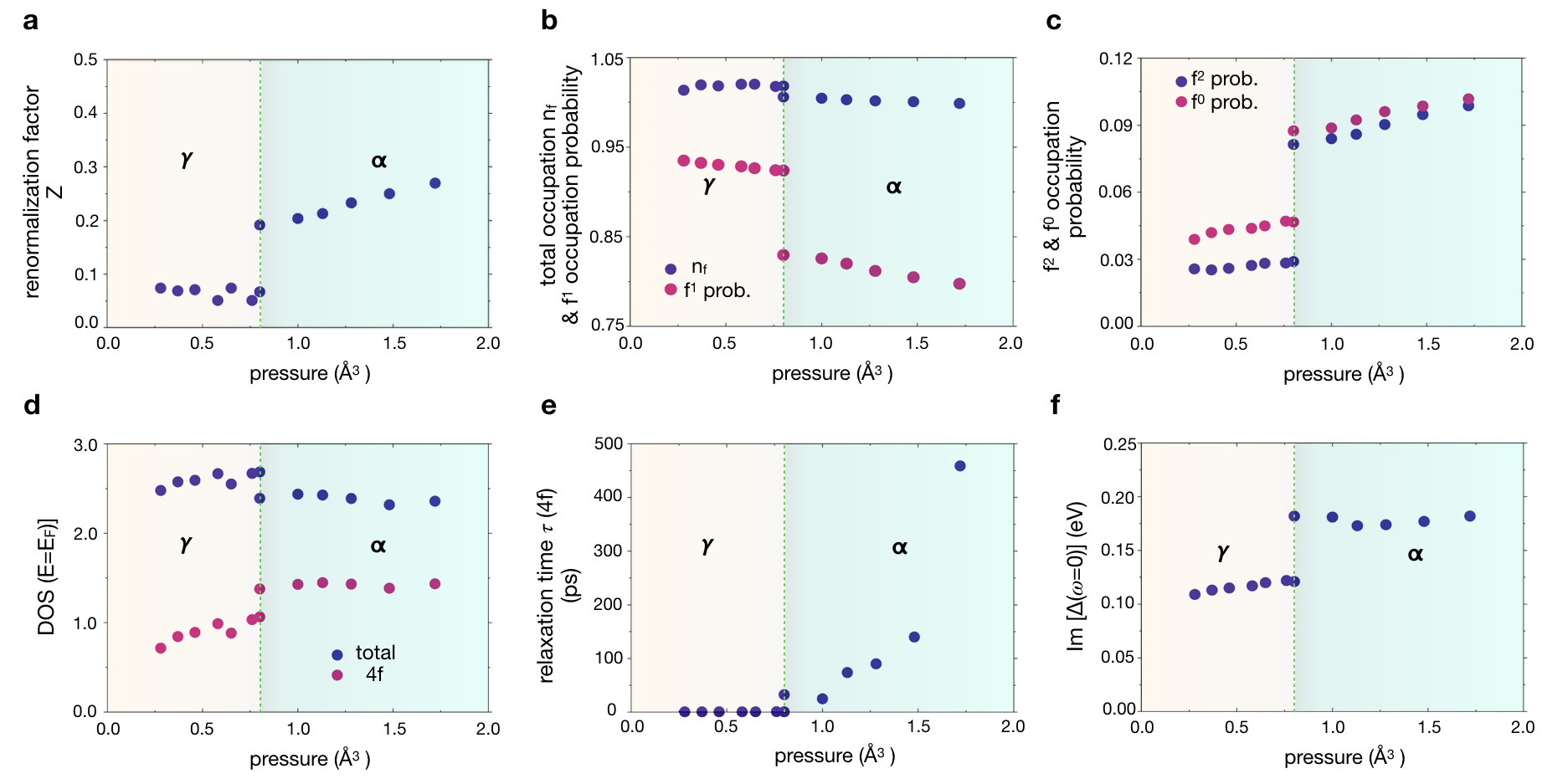} 
\caption{
(Color online) 
Pressure dependent DMFT physical parameters 
along the pressure-volume isotherm process at $T$ = 293K.
{\bf (a)} Renormalization factor $Z$.
{\bf (b)} Occupation probabilities of $f^{0}$ and $f{^2}$ configurations.
{\bf (c)} Occupation of $f$-electrons, $f^{tot}$, and 
the occupation probability of $f{^1}$ configuration.
{\bf (d)} Total and $4f$ DOS at {\EF}.
{\bf (e)} Relaxation time ${\tau}$
$( {\tau}^{-1}= -2 Im {\Sigma} )$.
{\bf (f)} Imaginary part of the DMFT hybridization function at {\EF}. 
}
\label{pressure_dep_DMFT}
\end{center}
\end{figure}

\begin{figure}[t]
\begin{center}
\includegraphics[width=0.99\textwidth] {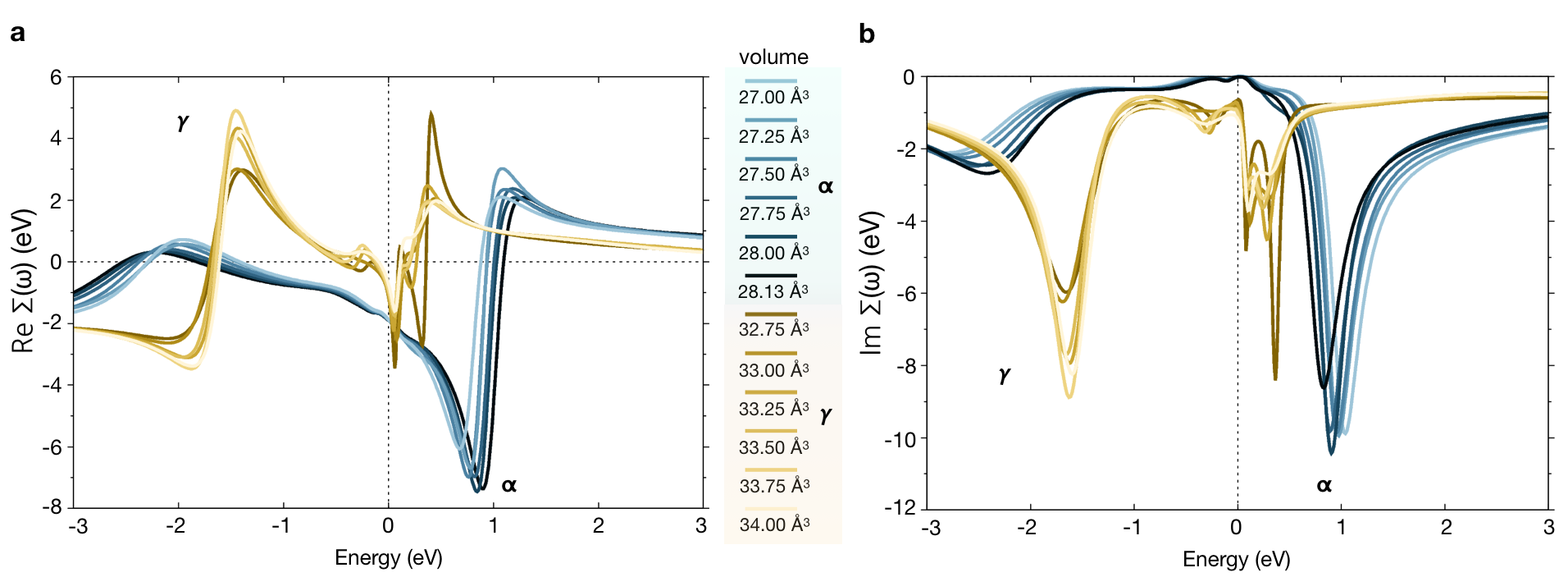} 
\caption{
(Color online) 
The self-energy $\Sigma(\omega)$
of single-particle Green's function of $4f_{j=5/2}$ orbital 
and its evolution
in the pressure-volume isotherm ($T$ = 293 K) process.
{\bf (a)} The evolution of the real part of $\Sigma(\omega)$
for the ${\gamma}$- and ${\a}$-phases. 
{\bf (b)} The evolution of the imaginary part of $\Sigma(\omega)$
for the ${\gamma}$- and ${\a}$-phases.
Note the qualitatively different behaviors of 
self-energies between the two phases.
}
\label{self_energy}
\end{center}
\end{figure}

\begin{figure}[t]
\begin{center}
\includegraphics[width=0.95\textwidth] {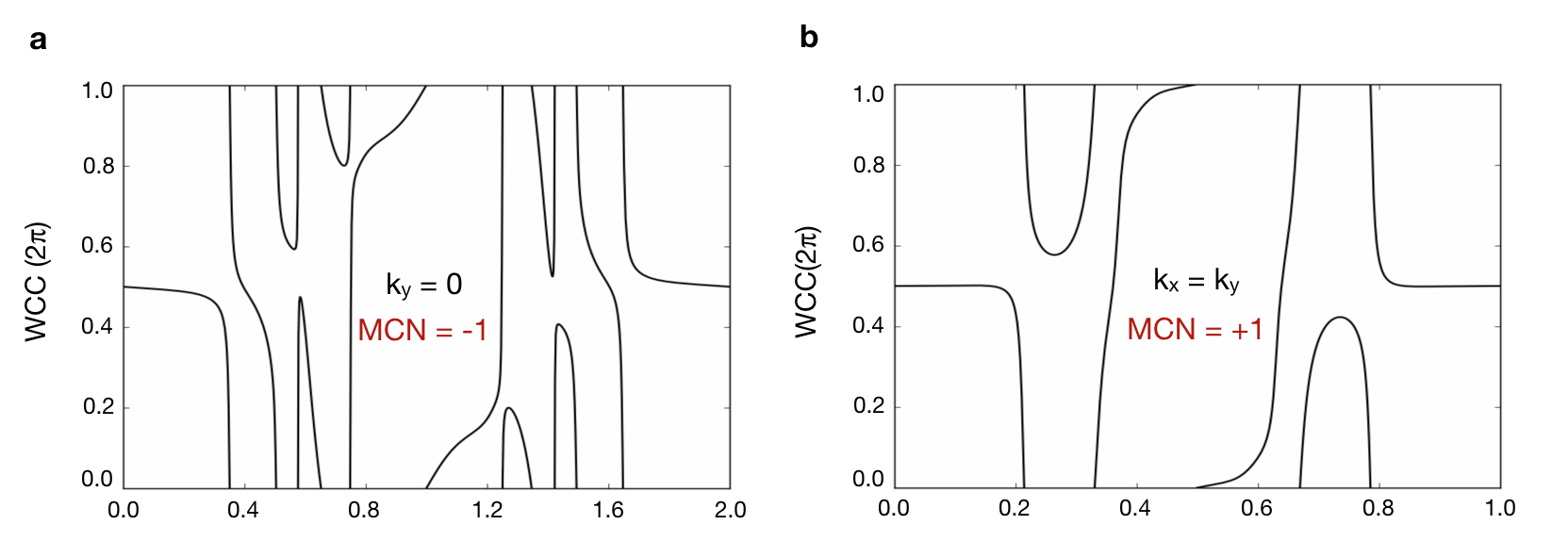} 
\caption{ 
The evolution of the Wannier charge center (WCC). 
Two distinct mirror Chern numbers (MCNs) for ${\alpha}$-Ce
are given along with the WCC evolutions. 
{\bf (a),(b)} MCNs for ${\alpha}$-Ce:
$-1$ for $k{_y}$ = 0 planes, and $+1$ for $k{_x}$ = $k{_y}$ planes, 
respectively.
These MCNs correspond to those with 
mirror eigenvalue of $+i$ and are obtained from the Wilson-loop method.
}
\label{WCC}
\end{center}
\end{figure}

\begin{figure}[t]
\begin{center}
\includegraphics[width=0.9\textwidth] {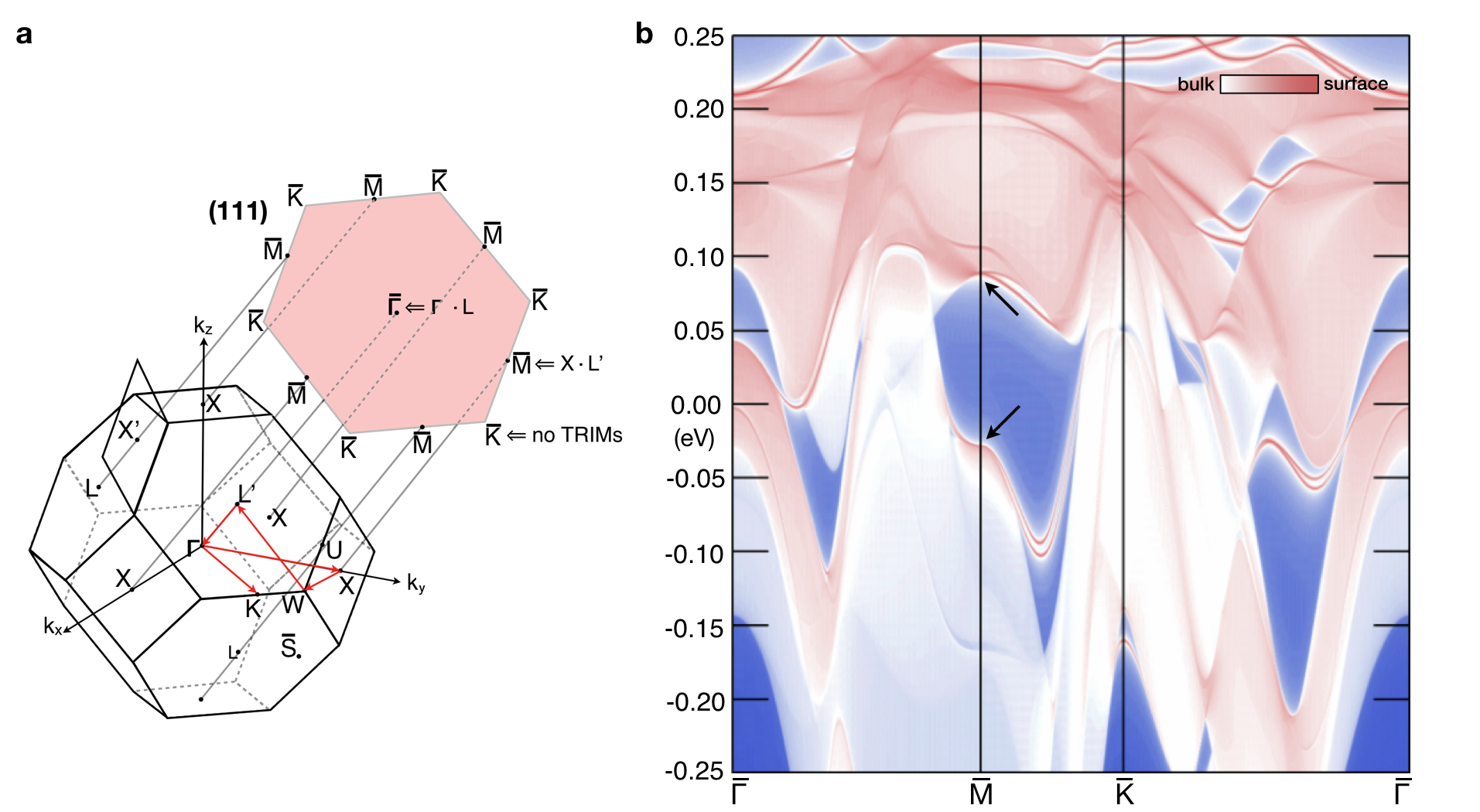} 
\caption{
(Color online) 
({\bf a}) 
Bulk and (111) surface BZ of fcc  ${\alpha}$-Ce.
The TRIM point $X$ in the bulk BZ is projected 
onto $\bar{M}$ in the surface BZ. 
({\bf b}) The (111) surface electronic structure of ${\alpha}$-Ce. 
It is calculated by a TB model with the semi-infinite slabs 
constructed from an effective topological Hamiltonian including 
Ce $6s$, $5p$, $d(t_{2g})$ and $4f_{5/2}$, $4f_{7/2}$ states
based on the DFT result (rescaled by 1/2 at around {\EF}).
Two possible Dirac-cone candidates
are indicated by two black arrows at $\bar{M}$.
}
\label{TSS_111}
\end{center}
\end{figure}

\section{PHASE BOUNDARIES}

Although some of the phase boundaries are still under debate 
due to the experimental uncertainty, 
it is generally accepted that there exist at least 
seven allotropic phases in solid Ce.  
In Fig. 1 of the text, fcc, bcc, and bct represent 
the face-centered cubic,
body-centered cubic, and body-centered tetragonal structures,
respectively.
${\beta}$-Ce has the double hexagonal-close packed (dhcp) structure.
 Among those phases, ${\gamma}$-${\alpha}$ phase boundary 
is of great interest.
The ${\gamma}$-${\alpha}$ phase transition was
discovered by Bridgman in 1927 \cite{a-g_discovery}.
But it became famous due to the isostructureness across the transition, 
which was observed in the x-ray experiment in 1940
\cite{isostructural_trans},
leading to a large volume decrease by 15\% in its maximum.
It is known that the size of isostructural volume collapse decreases
as the temperature ($T$) increases, 
and finally disappears at a critical point (CP).
According to recent high-quality X-ray diffraction 
data \cite{X-ray_diffraction}, 
the position of CP is located at 
the pressure ($P$) of 1.5 $\pm$ 0.1 GPa and
$T$ = 480 $\pm$ 10 K.
For the clear comparison of the electronic structures
between ${\alpha}$ and ${\gamma}$ phases,
we chose two representative points,
where both phases crystallize in face-centered cubic (fcc) structure
with the space group of \textit{Fm}$\bar{3}$\textit{m} (No.225),
and ${\alpha}$-Ce and ${\gamma}$-Ce have the lattice constants 
of 4.82 {\AA} at $P$ = 0.81 GPa and $T$ = 100 K (``red star")
and 5.16 {\AA} at the ambient pressure and  $T$ = 500 K 
(``blue star") in Fig. 1,
respectively \cite{a-Ce,g-Ce}. 
Both of them are relatively far from the CP.

\section{Lifshitz transition 
across the ${\gamma}$-${\alpha}$ transition}

In 1960, Lifshitz reported anomalies of 
thermodynamic and kinetic quantities 
due to an ``electron transition" 
via the variation of the topology of Fermi surfaces \cite{Lifshitz}.
Since ${\gamma}$-${\alpha}$ phase transition happens 
in the absence of structural symmetry change,
it is a good test material to observe the electron transition 
and corresponding anomalies in the thermodynamic and kinetic quantities.
Interestingly,
the size of isostructural volume collapse decreases
as $T$ increases.
In Fig. \ref{FS_compare}. as an example of the Lifshitz transition,
we have compared the Fermi surfaces (FSs) at two different volumes 
(i) $V$ = 34 \AA$^3$ ($T$ = 500 K, ambient pressure) 
where ${\gamma}$ phase is thermodynamically stable
and (ii) $V$ = 27.76 \AA$^3$ ($T$ = 100 K, $P$ = 0.88 GPa)
where ${\a}$ phase is thermodynamically stable, 
both of which are far from the phase boundary and 
the CP($P$ = 1.5 GPa). 
The FS topology in the low pressure ($V$ = 34 \AA$^3$) ${\gamma}$ phase, 
which arises mostly from the delocalized $5d$ electrons, 
is transformed drastically into a new topology
in the high pressure ($V$ = 27.76 \AA$^3$) ${\alpha}$ phase, 
as the coherent $4f$ band dominates the low energy spectrum.
In the ${\gamma}$ phase, the DMFT FS 
in Fig. \ref{FS_compare}{\bf a} 
shows a very similar FS to that of ``4f opencore" 
in Fig. \ref{FS_compare}{\bf c}.
Since the ``4f opencore" calculation excludes 
the $4f$ electron in the valence state, 
the great similarity between the DMFT FS and the ``4f opencore" FS
implies that the  $4f$ electrons in the ${\gamma}$ phase 
have almost localized nature.
On the other hand, in the ${\alpha}$ phase, 
the DMFT FS in Fig. \ref{FS_compare}{\bf d} 
has a good agreement with the DFT FS in Fig. \ref{FS_compare}{\bf e}, 
which indicates that the $4f$ electrons in the  ${\a}$ phase
behave as having a coherent quasi-particle band nature. 
The evolution from the localized to coherent band nature of $4f$ electrons 
is also shown in the volume-dependent
band structures in Fig. \ref{gap_feature}
along the pressure-volume isotherm at a given $T$ = 293 K.

Another interesting point is that Ce also
shows a Lifshitz-type transition only by $T$ change.  
A $T$-induced Lifshitz transition in the absence of 
a structural or magnetic phase transition is extremely rare
in weakly correlated systems.
This is because
the electronic structure does not change much by $T$ change.
In strongly correlated systems, however,
$T$ could be one of the sources
to drive the Lifshitz transition, as in the case of Ce,
because the electronic structure becomes
sensitive to the $T$ change.
Similar $T$-induced Lifshitz transitions were observed 
in the previous studies on CeIrIn$_5$ \cite{T-dep_Lifshitz1}
and golden-phase SmS \cite{T-dep_Lifshitz2}.

\section{Pressure-dependent evolutions of various DMFT Physical quantities}

Renormalization factor $Z$ is displayed 
in Fig. \ref{pressure_dep_DMFT}(a) 
as a function of pressure. 
Since the DMFT $f$-$f$ hopping strength and 
the DMFT $f$-$d$ hybridization function $\Delta(\omega)$
are effectively proportional to the $Z$ factor,
the enhancement of $Z$ factor helps the $f$ electrons 
to form a quasi-particle band around {\EF} 
in the ${\a}$-phase.
Figure \ref{pressure_dep_DMFT}(B) shows that
the total $f$ electron occupation
drops only slightly across the transition.
But, as shown in Fig. \ref{pressure_dep_DMFT}(b) and (c),
the abrupt reduction of
$f^1$ probability accompanied by the abrupt 
enhancement of
$f^0$ and $f^2$ probabilities indicates
that $f$ electrons in  ${\a}$-Ce become more or less delocalized.
Figure \ref{pressure_dep_DMFT}(D)-(F)
show the total and $4f$ DOS, the relaxation time ${\tau}$,
and the imaginary part of the DMFT hybridization function at {\EF},
$Im\Delta(0)$, across the ${\gamma}$-${\alpha}$ transition.
The relaxation time ${\tau}$ 
is obtained from the inverse of the imaginary part 
of the self-energy $\Sigma$. 
The contribution of $f$ electrons to the relaxation time 
is much bigger in ${\a}$-Ce by two orders of magnitude. 
But, in reality, ${\gamma}$-phase has 5d band at {\EF},
and so the experimental relaxation times of both phases 
are comparable \cite{opt_spec_Ce}.   
It is seen that $Im\Delta(0)$ increases by over 50 percent in average
across the transition.  This enhanced hybridization strength 
is one of the key ingredients 
that makes the system have non-trivial Z$_2$ topology. 
Figure \ref{self_energy} shows 
the self-energy $\Sigma(\omega)$ of $4f_{j=5/2}$ orbital 
and its evolution with respect to the pressure change.
The self-energy obtained on the imaginary frequencies
are transformed into that on the real frequencies
by using the maximum entropy method based on the analytic continuation.
The qualitatively different behaviors of real and imaginary self-energies 
between the ${\a}$ and ${\gamma}$ phases are remarkable,
which produce quite distinct electronic structures
between two.

\section{Topological Crystalline Insulator (TCI)-Type Nature}

Band inversion in combination with additional crystal symmetry
can lead to another type of topological insulator (TI) nature, 
the so-called, topological crystalline insulator (TCI) nature. 
One of the simple crystal symmetries
that is realized in real materials is the mirror symmetry.
When Bloch states in a system are symmetric under the mirror operation, 
then, mirror Chern numbers (MCNs) are useful topological invariants
to clarify the TI-type nature of materials.
We have used the Wilson-loop method 
to calculate the MCNs of ${\alpha}$-Ce.
Figure \ref{WCC} presents the MCN sets 
of the Bloch states on the corresponding 
mirror-symmetry planes with the mirror eigenvalue of $+i$ 
for ${\alpha}$-Ce.
The MCNs are obtained for the two mirror planes 
$k_{y}=0$ and $k_{x}=k_{y}$.
As shown in Fig. \ref{WCC} 
${\a}$-Ce has the nontrivial MCNs of  
(C$^{+i}_{k_{y}=0}$ = $-1$, C$^{+i}_{k_{x}=k_{y}}$ = $+1$), 
which are the same as those of g-SmS \cite{Ss2}.
This indicates that both systems would be TCI systems.

\section{Surface states at (111) surfaces}

For the (111) surface Brillouin zone (BZ), 
one bulk $X$  time-reversal invariant momentum (TRIM)
point is projected onto $\bar{M}$.
As shown in Fig. \ref{TSS_111},
in this case too, most surface states are buried
under the bulk-projected bands. 
But Dirac cone-like bands are vaguely seen at
around +90 meV  and $-30$ meV at $\bar{M}$,
as designated by black arrows.
